\documentclass[journal]{IEEEtran}
\ifCLASSINFOpdf

\else

\fi

\usepackage{cite}
\usepackage{amsmath,amssymb,amsfonts}
\usepackage{algorithmic}
\usepackage{graphicx,,subfig}
\usepackage{textcomp}
\usepackage{xcolor}
\usepackage[switch]{lineno}
\usepackage[ruled,vlined]{algorithm2e}
\usepackage{float}
\newcommand{\norm}[1]{\left\lVert#1\right\rVert}

\begin{document}

 \title{Multimodal Unrolled Robust PCA for Background Foreground Separation}

\author{Spencer Markowitz,
        Corey Snyder,
        Yonina C. Eldar, Minh N. Do
\thanks{The authors would like to thank Sandia National Laboratory (LDRD Project) and Texas Instruments for their support for this work.}}


\maketitle

\begin{abstract}
Background foreground separation (BFS) is a popular computer vision problem where dynamic foreground objects are separated from the static background of a scene.  Typically, this is performed using consumer cameras because of their low cost, human interpretability, and high resolution. Yet, cameras and the BFS algorithms that process their data have common failure modes due to lighting changes, highly reflective surfaces, and occlusion. One solution is to incorporate an additional sensor modality that provides robustness to such failure modes. In this paper, we explore the ability of a cost-effective radar system to augment the popular Robust PCA technique for BFS. We apply the emerging technique of algorithm unrolling to yield real-time computation, feedforward inference, and strong generalization in comparison with traditional deep learning methods. We benchmark on the RaDICaL dataset to demonstrate both quantitative improvements of incorporating radar data and qualitative improvements that confirm robustness to common failure modes of image-based methods.
\end{abstract}


%
%
%
%

\section{Introduction}

\IEEEPARstart{B}{ackground} foreground separation (BFS) is a fundamental task for many computer vision algorithms where dynamic foreground components are separated from the static background of a given scene. Successful BFS enables applications in intelligent surveillance such as vehicular traffic monitoring, industrial manufacturing, and human activity recognition \cite{bfs-applications}. A wide variety of approaches to BFS exist in the literature. Subspace methods and deep learning have emerged as the dominant techniques in the past decade due to their superior performance on popular benchmark datasets \cite{rpca-survey,deep-learning-survey}. 
Other techniques include statistical methods, fuzzy models, and cluster models. For a recent review, see \cite{classical-survey}.

\begin{figure}[t]
    \centering
    \includegraphics[width=\linewidth]{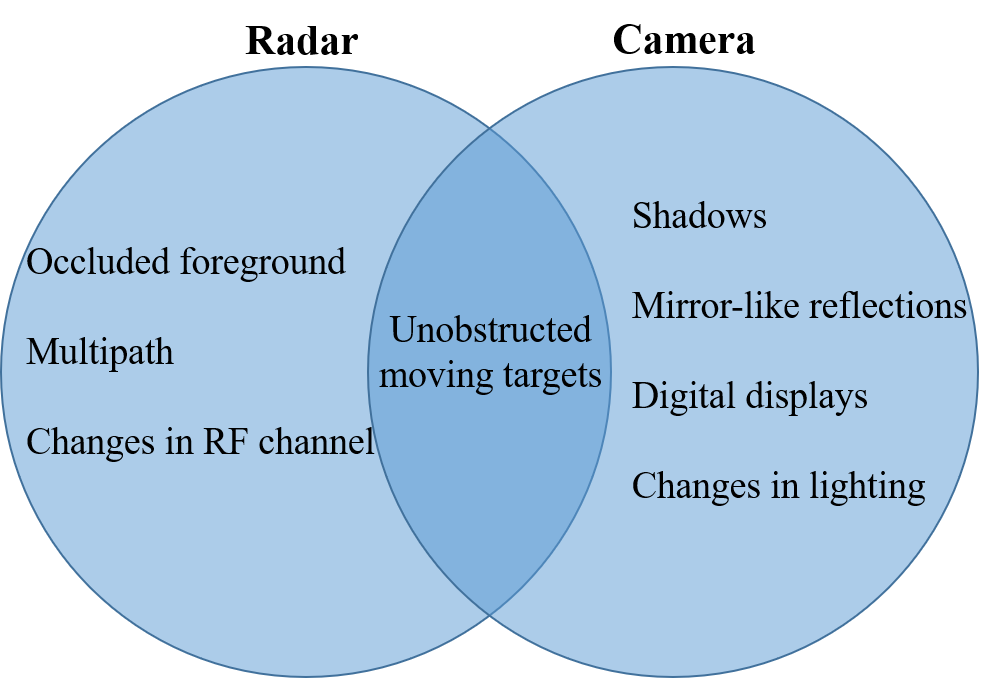}
    \caption{Examples of targets or phenomena that may potentially be detected as foreground for radar and camera sensors.}
    \label{fig:venn}
\end{figure}

Subspace methods seek to decompose an image sequence into the sum of a low-rank background and sparse foreground. Robust PCA (RPCA) \cite{rpca} is one highly influential method which solves the convex Principle Component Pursuit (PCP) program to perform this separation. While effective in many settings, PCP can take hundreds of iterations to converge using popular solvers such as ISTA or ADMM, and subspaces must be re-computed when new data is made available. These shortcomings have been explored with faster optimization algorithms \cite{faster-rpca}, real-time RPCA under a correlated sparse outliers assumption \cite{realtime-rpca}, and online versions of RPCA \cite{orpca}.

In addition to these drawbacks, subspace methods are also sensitive to changes in lighting, camera alignment, and dynamic backgrounds like moving water and trees. For example, a small translation of the camera defines a completely new low-rank subspace for the background. There is no feature learning in subspace methods, thus such subtle changes cannot be recognized or properly discarded without explicitly imposing greater structure on the subspace model. Supervised deep learning techniques \cite{deep-learning-survey} have been shown to address these limitations and even provide human-level performance on supervised learning benchmarks like CDnet14\cite{cdnet14} and Scene Background Initialization 2015 \cite{SBI2015}. With hand-labeled ground-truth examples, Convolutional Neural Networks (CNNs) learn rich features that focus on salient changes in a scene and are robust to changes like the aforementioned camera shift due to the translation-invariance of the convolution operator. Deep learning algorithms often require an expensive fitting or training process like subspace methods; however, they are able to be deployed immediately on unseen data without the re-fitting that subspace methods require. Example models include FgSegNet \cite{fgsegnet} and CascadeCNN \cite{cascade-cnn}.

\begin{figure*}[ht]
\centering
    \subfloat[\centering RGB Image]{{\includegraphics[width=.33\linewidth]{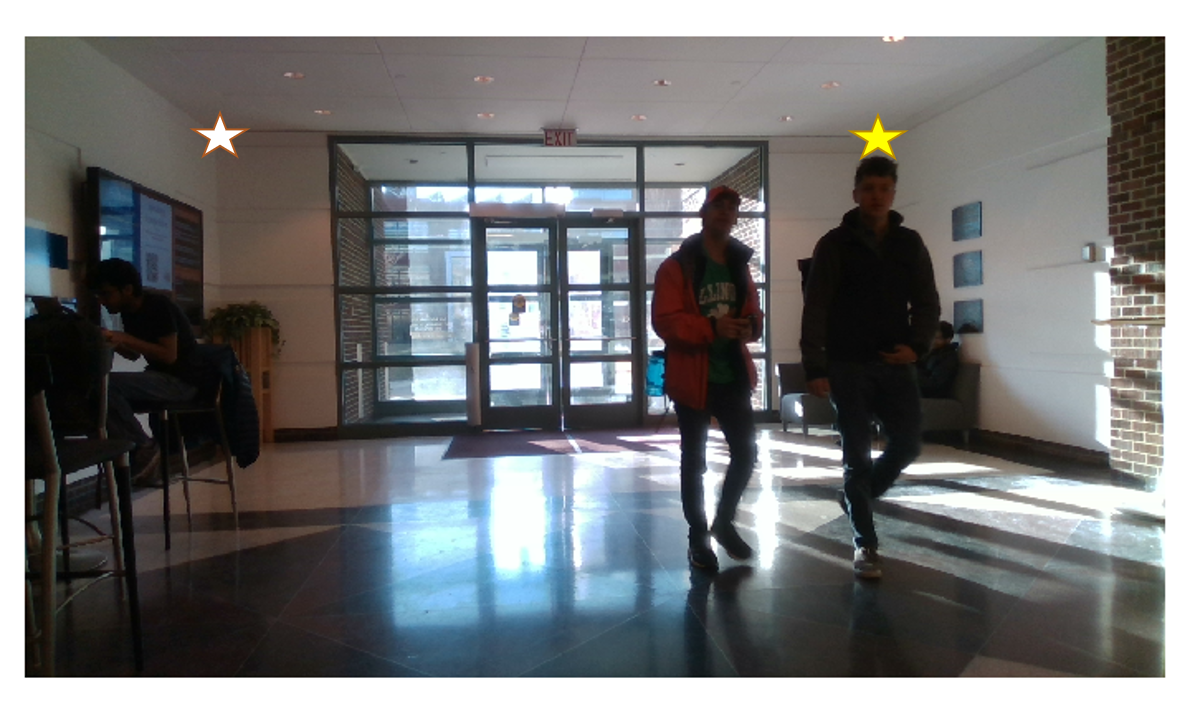} }}%
    \subfloat[\centering Range-azimuth heatmap with clutter ]{{\includegraphics[width=.33\linewidth, trim=55 145 48 170, clip]{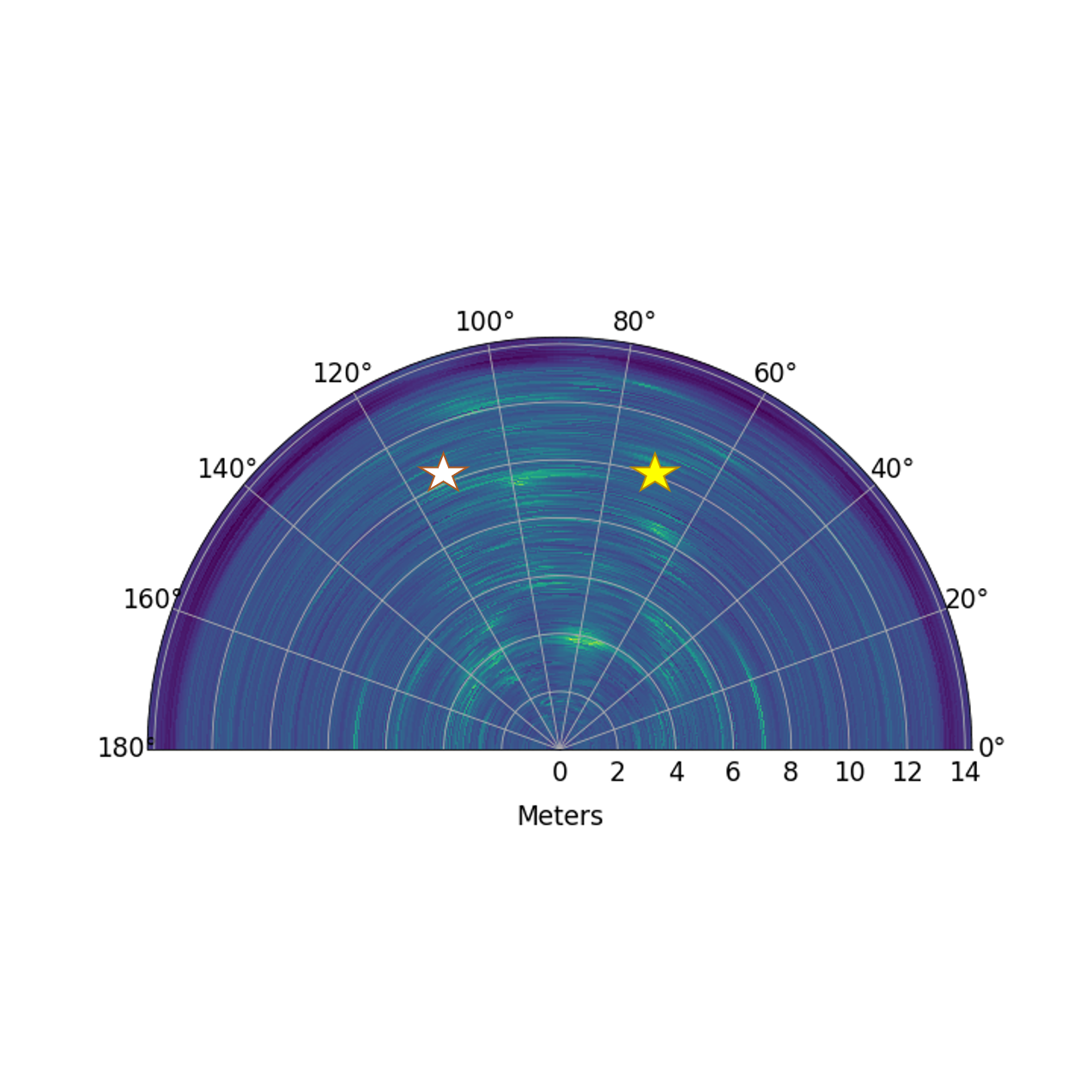} }}%
    \subfloat[\centering Range-azimuth heatmap without clutter ]{{\includegraphics[width=.33\linewidth, trim=30 80 25 90, clip]{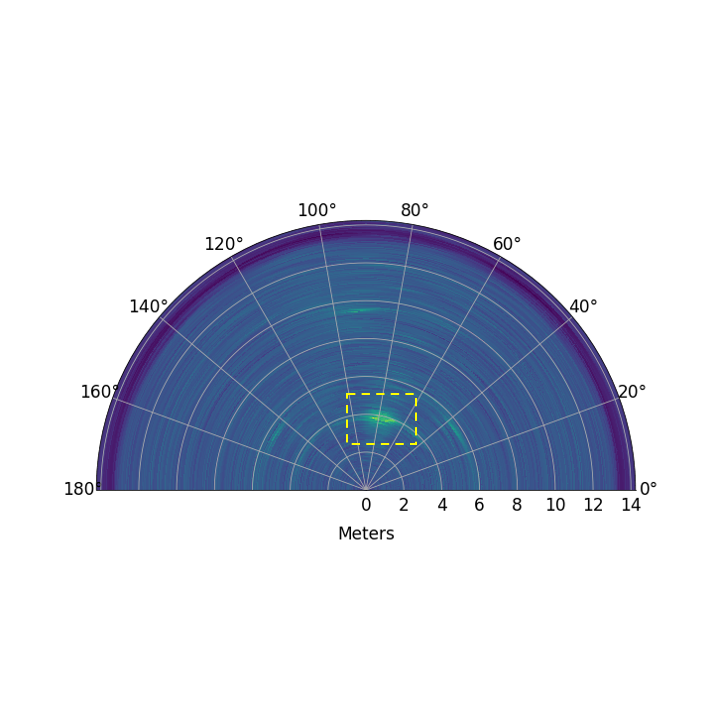} }}%
    
    \caption{Example range-azimuth heatmaps for a particular scene (scene B in the appendix). As a reference, there are two stars at the back corners of the lobby in both (a) and (b). Notice in (b) the wall is not represented by a continuous line of reflections. Figure (c) shows the resulting clutter-free heatmap using the methods described in Sec.~\ref{sec:radar_proc}. With the clutter suppression, the humans (highlighted by the bounding box) become much more visible within the heatmap.}%
    \label{fig:radar_example}
\end{figure*}

The key shortcoming of deep learning methods is that they require expensive pixel-level ground truth for hundreds of images. These deep CNNs typically have on the order of millions of learnable parameters. Thus, when limited ground truth is available, they are prone to overfitting and poor generalization to unseen data. Unsupervised deep learning approaches have been suggested; however, they lag behind their supervised counterparts \cite{deep-context-prediction}.

In this work, we operate in the unsupervised BFS problem setting where no hand-labeled data is available during training and side information from an additional sensing modality is used alongside camera data. To the best of our knowledge, this is the first such work in BFS that incorporates radar sensing into a BFS algorithm. Automotive/consumer radar sensing has recently seen a great deal of interest due to its affordability, compact size, and ability to sense in conditions where cameras perform poorly.  In the context of BFS, radar does not detect some undesirable foreground components that cameras capture. Examples include shadows, changes in lighting, reflections and digital screens as depicted in the Venn diagram in Fig.~\ref{fig:venn}. Moreover, compared to its camera counterpart, detecting moving targets in radar data is much simpler. This is achieved by measuring subtle changes in the phase of the received signal. This allows radar to detect salient motion in as little as one frame depending on the type of radar. A comparison between camera and radar data of the same scene is shown in Fig.~\ref{fig:radar_example}. Unlike cameras, radar sensing's shortcomings include low angular resolution, specularity, and multipath, the first two of which can be seen in Fig.~\ref{fig:radar_example}b.

In designing our algorithm we make the practical considerations for (1) real-time computation, (2) robustness to unseen data, and (3) cost-effectiveness. Towards the first two points, we leverage the advantages of both subspace and deep learning models via the emerging technique of algorithm unrolling. First proposed in \cite{first_unroll} for sparse coding, an iterative algorithm is \emph{unrolled} or \emph{unfolded} by representing the $k$'th iteration as the $k$'th layer in a feedforward network. The result of the $k$'th layer is fed as the input to layer $k+1$ where common operations of iterative algorithms such as shrinkage operators function as the non-linearities in traditional deep nets, e.g. ReLU. Unrolled networks have been shown to achieve the same performance as their iterative counterparts using dramatically fewer layers. This means real-time computation is possible on both seen and unseen data without sacrificing performance. Furthermore, compared to state of the art deep nets, unrolled neural networks often use far fewer parameters, require less training data, and maintain a high level of interpretability due to the structure imposed by its accompanying ``white box'' iterative algorithm \cite{monga2021algorithm}.

In addition to sparse coding, algorithm unrolling has also been used to tackle a wide variety of problems employing well studied algorithms and supplementing key assumptions with data-driven techniques. Examples of unrolling algorithms into feedforward networks include image deblurring \cite{image_deblur}, phase retrieval \cite{phase_retrieval}, channel estimation \cite{viterbi_net}, and clutter suppression in ultrasound \cite{corona}.

In this paper, we extend the work of CORONA \cite{corona}, which is an unrolled Robust PCA technique. We combine radar side-information with camera data into the Robust PCA objective to re-weight the penalty of the sparse foreground. We present the ISTA algorithm for this radar-modified objective and refer to this procedure as RISTA. We then unroll RISTA into a feedforward convolutional neural network we call Radar Unrolled Shrinking and Thresholding Incorporating Convolutions, or RUSTIC. To ensure our method is cost-effective and practical, we use frequency-modulated continuous wave (FMCW) radar for our experimentation. FMCW radar is low cost (the Texas Instruments IWR1443 we use costs \$12 USD) and small in size (can be put on a single PCB). FMCW radar can transmit \emph{and} receive simultaneously thus allowing detection of targets at very close range. We perform quantitative evaluation on the RaDICaL dataset \cite{radical} and demonstrate that RUSTIC delivers competitive and sometimes superior performance to its iterative counterpart on both seen and unseen data while enabling real-time computation. We also compare RUSTIC to the CORONA model that only utilizes camera data and a conventional deep learning segmentation model in the U-Net \cite{unet}. We show a clear improvement in quantitative performance by incorporating radar side-information for shallower unrolled models and demonstrate these unrolled models generalize far better to unseen scenes than the U-Net while using orders of magnitude fewer parameters. Furthermore, we provide qualitative examples illustrating the effectiveness of both the camera and radar modalities to correct errors from one another through the RUSTIC framework.

The rest of the paper is organized as follows. Section~\ref{sec:math_sec} defines our problem setting and motivates the incorporation of radar data into the RPCA objective to form our iterative RISTA algorithm. In Section~\ref{sec:nn}, we explain how RISTA is unrolled into our RUSTIC model. Section~\ref{sec:experiment} details our quantitative and qualitative experimental results for RUSTIC and related works on the RaDICaL dataset. Finally, we conclude in Section~\ref{sec:conclusion} and provide suggestions for future work using RUSTIC and sensor fusion in BFS.

\section{Sensor Fusion for BFS}
\label{sec:math_sec}
\subsection{Problem Setup}
We consider the scenario where $M$ radar frames and $M$ camera frames, each separated by $\Delta t$, observe the same scene and are synchronized in time. Using both the camera and radar data, we aim to separate the camera data into its background and foreground components. Let $\mathbf{D}_m\in \mathbb{R}^{H \times W}$ be a single frame in our video sequence for a given scene. For the radar data, we assume the radar transmits a constant amplitude sawtooth waveform such that for a given chirp interval, $0<t<T$, the frequency can be expressed as $f_c+Bt$ where $f_c$ is the starting frequency and $B$ is the chirp slope. This yields the following transmitted waveform
\begin{align}
    S_{tx} = A_{tx}\cos \left(2\pi\left(f_ct+\frac{B}{2}t^2 \right) \right)
\end{align}
where $A_{tx}$ is the signal amplitude. After the transmitted signal is reflected from a target back to the radar's receivers, the signal is subsequently mixed with the transmitted signal and low-pass filtered to obtain the intermediate frequency (IF) signal. The IF signal is then sampled $N_s$ times during the chirp's interval. We will also assume that $T\ll\Delta t$ allowing us to include multiple radar chirps in each radar frame. Accordingly, we will consider one complete radar frame to contain $N_c$ sequential chirps each received and sampled at $N_a$ receivers. This results in a single frame of radar data taking the form $\mathbf{R}_m\in \mathbb{C}^{N_s \times N_a \times N_c}$.

Next, we follow the RPCA \cite{rpca} subspace method for BFS closely and seek to separate our camera data $\mathbf{D}$ into its low-rank and sparse components, $\mathbf{L}$ and $\mathbf{S}$, respectively. We accomplish this by first vectorizing each frame of $\mathbf{D}$ such that $\mathbf{D}$, $\mathbf{L}$ and $\mathbf{S}$ all belong to $\mathbb{R}^{HW\times M}$. The low-rank+sparse decomposition objective is commonly stated as follows:

 \begin{equation}
     \min_{\mathbf{L}, \mathbf{S}}~\mathrm{rank}(\mathbf{L})+||\mathbf{S}||_0,~\mathrm{s.t.}~\mathbf{D}=\mathbf{L}+\mathbf{S}.
     \label{eqn:rpca-obj}
 \end{equation}
 Since this program is non-convex, we use the popular convex relaxation 

\begin{equation}
    \min_{\mathbf{L}, \mathbf{S}}~||\mathbf{L}||_*+\lambda||\mathbf{S}||_1,~\mathrm{s.t.}~\mathbf{D}=\mathbf{L}+\mathbf{S}
    \label{eqn:pcp-obj}
\end{equation}
where $||\cdot||_*$ and $||\cdot||_1$ are the nuclear and $l_1$ norms, respectively. In the following section, we describe how radar side information can be incorporated into (\ref{eqn:pcp-obj}) and then present the resulting iterative solver that will become the foundation for our unrolled feedforward network.

\subsection{Incorporating Radar}
\label{sec:motivation}

One of the many advantages that radar systems have over cameras is their ability to easily localize motion within a frame and remove any static clutter. Here, we perform this relatively simple operation first and then incorporate the clutter-free radar return in the BFS of the camera data. 
We assume that the clutter-free radar returns can provide useful information on where foreground is likely to exist in the camera data. Although the camera's low rank component is a function of its own sparse component, we do not use the radar data directly with the low rank prediction. We make this choice because the radar data does not provide informative cues like it can for the foreground component. For example, specularity and the physical properties of common construction materials can prevent portions of walls from being detected. These impacts are seen in Fig.~\ref{fig:radar_example}b where the walls have many undetected patches. Conversely, moving targets are consistently detected as depicted in Fig.~\ref{fig:radar_example}b and \ref{fig:radar_example}c after clutter suppression. Thus, we do not use the extracted static clutter and choose only to use the moving targets' radar reflections to convey the locations and associated likelihoods of foreground in the camera images. 

\begin{figure*}[ht]
    \includegraphics[width=\textwidth]{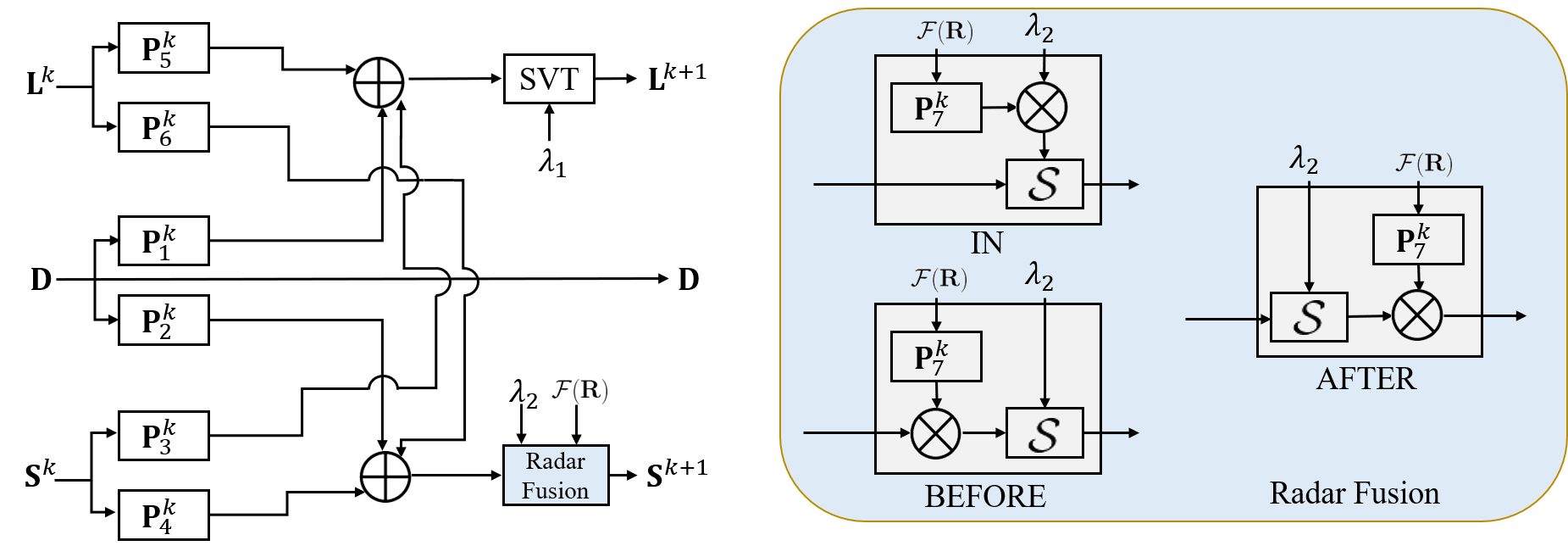}
    \caption{On the left is a depiction of a single layer of our RUSTIC architecture. On the right, we show the three options we consider for how to incorporate the radar into the unrolled network based on Alg.~\ref{alg:ista_mod}. While incorporating the radar \emph{in} the shrinkage operator is the most accurate interpretation of Alg.~\ref{alg:ista_mod}, we also experiment with two looser interpretations, namely using the radar \emph{before} and \emph{after} the shrinkage operator.}
    \label{fig:three_models}
\end{figure*}

Intuitively, we seek to modify the RPCA objective in (\ref{eqn:pcp-obj}) to make sparse foreground contributions less costly in regions where the clutter-free radar return is high and make contributions in regions where it is low more difficult to admit. We therefore modify (\ref{eqn:pcp-obj}) and suggest solving

\begin{equation}
    \min_{\mathbf{L},\mathbf{S}} ||\mathbf{L}||_* + \lambda ||\mathbf{S}\circ\mathcal{F}(\mathbf{R})||_1,~\mathrm{s.t.}~\mathbf{D}=\mathbf{L}+\mathbf{S}
    \label{eq:our-pcp-obj}
\end{equation}
where $\mathcal{F}(\cdot): \mathbb{C}^{M\times N_s \times N_a \times N_c} \mapsto \mathbb{R}^{HW\times M}$ maps the radar data to a clutter-free, real-valued weight matrix and $\circ$ is the element-wise Hadamard product. We will make the radar processing pipeline that forms $\mathcal{F}(\cdot)$ concrete in the following subsection.

The program in (\ref{eq:our-pcp-obj}) can be solved efficiently using a number of solvers such as ADMM or ISTA. We choose to closely follow the derivation presented in \cite{corona} and select ISTA with the addition of the radar side information. We introduce the equality constraint in (\ref{eqn:pcp-obj}) and (\ref{eq:our-pcp-obj}) from the objective function as a quadratic penalty and add measurement matrices $\{\mathbf{H}_i\}_{i=1}^2$ for the low-rank and sparse components. We also multiply the radar input with its own measurement matrix, $\mathbf{H}_3$ to account for proper scaling and filtering as it corresponds with the camera data. This results in the problem
\begin{equation}
    \min_{\mathbf{L},\mathbf{S}} 
    ||\mathbf{D}-\mathbf{H}_1\mathbf{L}-\mathbf{H}_2\mathbf{S}||_F^2 +
    \lambda_1 ||\mathbf{L}||_* + \lambda_2 ||\mathbf{S}\circ\mathbf{H}_3\mathcal{F}(\mathbf{R})||_1
    \label{eqn:our-ista-obj}
\end{equation}
where $\lambda_1, \lambda_2 >0$. We set $\lambda_1$ and $\lambda_2$ according to the conditions used in the original RPCA paper \cite{rpca}. The choice of measurement matrices $\{\mathbf{H}_i\}_{i=1}^3$ is application-dependent. We make the simplifying assumption that each measurement matrix is identity since we have no prior knowledge of a more informed choice. We still leave the operators in place since they will be further abstracted to enrich our unrolled model described in Section~\ref{sec:nn}.

Following \cite{corona}, the modified radar-ISTA, \emph{RISTA}, is shown in Alg.~\ref{alg:ista_mod} where $\mathbf{X}^H$ is the Hermitian transpose, $\mathbf{I}$ is an appropriately sized identity matrix, $\mathcal{S}_\tau(x) := \text{sgn}(x)\max(|x|-\tau,0)$, and $\mathrm{SVT}_\tau(\mathbf{X}) :=\mathbf{U}\mathcal{S}_\tau(\mathbf{\Sigma)V}^T $ where $\mathbf{X = U\Sigma V}^T$ is the singular value decomposition of $\mathbf{X}$. The constant $\mu$ represents the step size for the proximal gradient operator and is given by the spectral norm of $\mathbf{H}^H\mathbf{H}$ where
\begin{equation}
    \mathbf{H} = \begin{bmatrix}\mathbf{H}_1\\\mathbf{H}_2 \end{bmatrix}.
\end{equation}
 When computing $\mathbf{S}^{k+1}$ in Alg.~\ref{alg:ista_mod}, we use a different threshold in the shrinkage operator for each column based on the processed radar data $\mathcal{F}(\mathbf{R})$ along with the terms $\mu, \lambda_2$, and $\mathbf{H}_3$. As such, for frame $m$, row $h$, and column $w$ the threshold in the shrinkage operator can be taken as the $hW+w$'th entry in $\mu\lambda_2 \mathbf{H}_3 \mathcal{F}(\mathbf{R}_m)\in\mathbb{R}^{HW}$.
\begin{algorithm}
\SetAlgoLined
\textbf{Input:} $\mathbf{D},\mathcal{F}(\mathbf{R}),\lambda_1,\lambda_2>0$\\
\textbf{Output:} $\mathbf{L}^{K_\text{max}},\mathbf{S}^{K_\text{max}}$\\
\textbf{Initialize:} $\mathbf{S}^0=\mathbf{L}^0=\mathbf{0}, k = 0$\\
 \While{not converged or $k < K_\text{max}$}{
    $\mathbf{G}_1^k = \left( \mathbf{I} - \mu \mathbf{H}_1^H \mathbf{H}_1\right) \mathbf{L}^k  - \mathbf{H}_1^H \mathbf{H}_2 \mathbf{S}^k + \mathbf{H}_1^H \mathbf{D} $\\
    
    $\mathbf{G}_2^k = \left( \mathbf{I} - \mu \mathbf{H}_2^H \mathbf{H}_2\right) \mathbf{S}^k  - \mathbf{H}_2^H \mathbf{H}_1 \mathbf{L}^k + \mathbf{H}_2^H \mathbf{D} $\\
  
    $\mathbf{L}^{k+1} = \text{SVT}_{\mu\lambda_1}(\mathbf{G}_1^k)$\\
    $\mathbf{S}^{k+1} = \mathcal{S}_{\mu\lambda_2 \mathbf{H}_3 \mathcal{F}(\mathbf{R})}(\mathbf{G}_2^k)$\\
    $k \leftarrow k+1$
   }
\caption{RISTA for minimizing (\ref{eqn:our-ista-obj})}
\label{alg:ista_mod}
\end{algorithm}

\subsection{Radar Processing}
\label{sec:radar_proc}
As mentioned in Section \ref{sec:motivation}, radars allow for easy clutter suppression compared to cameras. Unlike cameras, clutter suppression for radar only requires data from a single radar frame and, as a result, each radar frame can be processed independently. We will drop the $m$ subscript and let $\mathbf{R:=R}_m\in\mathbb{C}^{N_s\times N_a \times N_c}$ for notational brevity in this section. 

The first step in processing the raw samples from the radar is to compute the range of the reflections in each chirp from each antenna \cite{radical}. This is performed by taking the Fast Fourier Transform (FFT) along each chirp's IF signal, or along the first dimension of $\mathbf{R}$ such that 
\begin{align}
    \hat{\mathbf{R}}_{k_s n_a n_c} = 
    \text{FFT}_{n_s}\{\mathbf{R}_{n_s n_a n_c}\}
\end{align}
and $k_s$ gives our frequency domain index. After the range information is computed, we remove the clutter by computing the mean across the chirps within the frame and then subtract it from the range information data
\begin{align}
    \mathbf{\mu}_{k_s n_a} &= \frac{1}{N_c} \sum^{N_c}_{n_c=1} 
    \hat{\mathbf{R}}_{k_s n_a n_c}\\
    \Tilde{\mathbf{R}}_{k_s n_a n_c} &= \hat{\mathbf{R}}_{k_s n_a n_c} - \mathbf{\mu}_{k_s n_a}.
\end{align}

Once the static clutter is removed, we use the multiple antennas to determine the received power at each bearing. The data used for our experiments only recorded data using a 1D uniform linear array with resolution along the azimuthal axis because of the low resolution in the elevation axis.
\begin{figure}[h!]
    \centering
    \includegraphics[width=.8\linewidth]{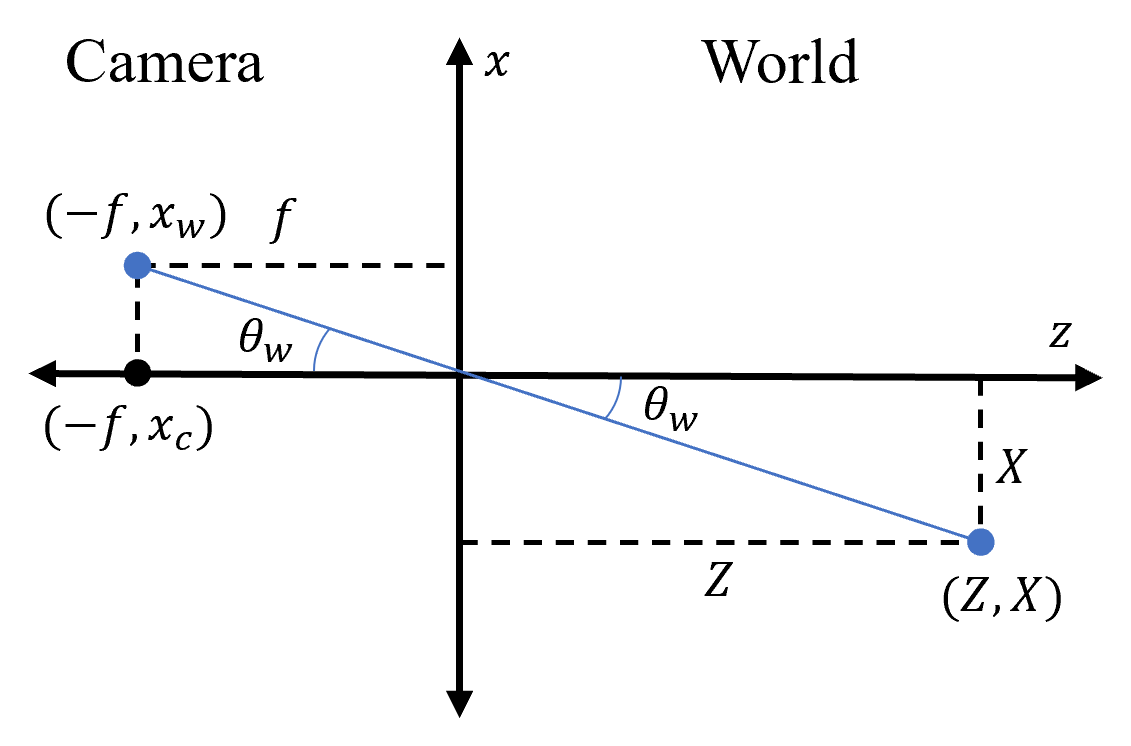}
    \caption{Depiction of a pinhole camera model.}
    \label{fig:pinhole}
\end{figure}
We use the standard pinhole camera model with focal length $f$ and camera center column $x_c$  e.g. $x_c = x_{W/2}$ to map a point in space to its horizontal pixel coordinates. As such, a point with horizontal coordinate $X$ and depth $Z$ is mapped to 
\begin{equation}
    (X,Z) \mapsto x_w := \frac{fX}{Z}+x_c
\end{equation}
as shown in Fig. \ref{fig:pinhole}.
Since our data assumes the radar and camera are coplanar and stacked vertically on top of each other, we may compute the bearing $\theta_w$ corresponding to each of the image's columns $x_w$ for $w = 1, ..., W$ as
\begin{align}
    \theta_w = \arctan\left(\frac{x_w-x_c}{f} \right).
    \label{eqn:bearing-eqn}
\end{align}

To compute the power at each $\theta_w$, we use Minimum Variance Distortionless Response beamforming \cite{mvdr}. This is accomplished by first computing the covariance matrix $\mathbf{\Sigma}_{k_s}$ for each range slice $\Tilde{\mathbf{R}}_{k_s}$ such that 
\begin{align}
    \mathbf{\Sigma}_{k_s} = \frac{1}{N_c}\Tilde{\mathbf{R}}_{k_s} \Tilde{\mathbf{R}}_{k_s}^H.
\end{align}
The received power at each range and angle is computed as
\begin{align}
    \mathbf{P}(k_s,\theta_w) = \frac{1}{\mathbf{a}(\theta_w) \mathbf{\Sigma}_{k_s}^{-1} \mathbf{a}^H(\theta_w)}
\label{eqn:mvdr}
\end{align}
where the steering vector $\mathbf{a}(\theta_w)$ simplifies to $[1, e^{-j\pi\sin(\theta_w)},...,e^{-j(N_a-1)\pi\sin(\theta_w)}]$ because the antennas are spaced half a wavelength apart.
We then take the $\log(\cdot)$ in (\ref{eqn:mvdr}) since the data often spans many orders of magnitude. Finally, we sum over the range dimension because the camera data lacks any depth information: 
\begin{align}
    \mathbf{P}(\theta_w) = \sum_{k_s = 1}^{N_s} \log[\mathbf{P}(k_s,\theta_w)].
\end{align}

We use the associated bearing $\theta_w$ for each column in the image data according to (\ref{eqn:bearing-eqn}) to compute the received power at each column and form $\mathbf{P}\in\mathbb{R}^W$.  Although $\mathbf{P}$ has no elevation data and can therefore be expressed as a 1D vector, we assert its size to be the same as each camera image so it can be multiplied elementwise with the camera data in Alg.~\ref{alg:ista_mod}. Thus, we expand $\mathbf{P}$ (with a slight abuse of notation) to be of shape $(H, W)$ by making each row identical. To summarize, $\mathbf{P}$ represents the result of $\mathcal{F}(\mathbf{R})$ that is computed independently for each radar frame $m\in[M]$.

\section{Unrolled Network with Radar}
\label{sec:nn}

\subsection{Model Architecture}
An iterative algorithm can be modeled as an unrolled neural network where the $k$'th layer corresponds to the $k$'th iteration \cite{sprechmann2015learning,monga2021algorithm}. As in \cite{corona}, we replace matrix multiplication using $\mathbf{H}_{\{1,2\}}$ with 2D convolutional layers $\{\mathbf{P}_i^k\}_{i=1}^6$ as well as multiplication with $\mathbf{H}_3$ with 1D convolution layers $\mathbf{P}_7^k$, and learn $\lambda_i^k$ for the shrinkage and SVT operations. The choice of 2D and 1D convolutional operators (as opposed to fully connected layers) promotes spatial coherence, reduces the number of learnable parameters, and provides the network with the desirable property of translation invariance. All together, Alg.~\ref{alg:ista_mod} can be represented as a multi-layer feedforward network with each layer being described by
\begin{align}
\begin{split}
    \mathbf{L}^{k+1} &= \text{SVT}_{\lambda_1^k}\{\mathbf{P}^k_5 * \mathbf{L}^{k} + \mathbf{P}^k_3 * \mathbf{S}^{k} + \mathbf{P}^k_1 * \mathbf{D}\} \\
    \mathbf{S}^{k+1} &= \mathcal{S}_{\lambda_2^k \mathbf{P}^k_7 * \mathcal{F}(\mathbf{R}) }\{\mathbf{P}^k_6 * \mathbf{L}^{k} + \mathbf{P}^k_4 * \mathbf{S}^{k} + \mathbf{P}^k_2 * \mathbf{D}\}
\end{split}
\label{eqn:in_description}
\end{align}
with $*$ being the convolution operator and $\mathbf{S}^0=\mathbf{L}^0=\mathbf{0}$. The shrinkage operator here follows the same notation as in Alg.~\ref{alg:ista_mod} where the threshold for layer $k$, frame $m$, row $h$, and column $w$ is determined by the $hW+w$'th entry in $\lambda_2^k \mathbf{P}^k_7 * \mathcal{F}(\mathbf{R}_m)$. The image data, $\mathbf{D}$, $\mathbf{L}$, and $\mathbf{S}$, are of shape ($M, H, W$). In order to perform the SVT operation, we vectorize the result of the convolution and addition operations for the updated low-rank component and stack along the second dimension to yield a shape of $(HW, M)$. We undo this procedure after SVT is performed. It is important to note here that while the iterative algorithm uses the same thresholds and measurement matrices for all iterations, the unrolled model learns different filters and thresholds for each layer. This is a unique advantage of unrolled networks with respect to their iterative counterparts.

We also experiment with looser interpretations of Alg.~\ref{alg:ista_mod} where instead of incorporating the radar directly \emph{in} the sparse shrinkage operator as in (\ref{eqn:in_description}), we incorporate it \emph{before} or \emph{after} as described in (\ref{eqn:before_description}) and (\ref{eqn:after_description}), respectively, and shown in Fig.~\ref{fig:three_models}:
\begin{equation}
    \mathbf{S}^{k+1} = \mathcal{S}_{\lambda_2^k}\{(\mathbf{P}^k_7 * \mathcal{F}(\mathbf{R})) \circ (\mathbf{P}^k_6 * \mathbf{L}^{k} + \mathbf{P}^k_4 * \mathbf{S}^{k} + \mathbf{P}^k_2 * \mathbf{D})\}
    \label{eqn:before_description}
\end{equation}
\begin{equation}
    \mathbf{S}^{k+1} = \mathcal{S}_{\lambda_2^k}\{\mathbf{P}^k_6 * \mathbf{L}^{k} + \mathbf{P}^k_4 * \mathbf{S}^{k} + \mathbf{P}^k_2 * \mathbf{D}\} \circ (\mathbf{P}^k_7 * \mathcal{F}(\mathbf{R})) 
\label{eqn:after_description}
\end{equation}
We designate these three variations (\emph{in}~(\ref{eqn:in_description}), \emph{before}~(\ref{eqn:before_description}), and \emph{after}~(\ref{eqn:after_description})) as Radar Unrolled Shrinking and Thresholding Incorporating Convolutions (RUSTIC). In practice, we find the model that incorporates radar \emph{after} the shrinkage operator performs the best. Section~\ref{sec:experiment} will compare these three variations.

\subsection{Model Training}
To train the RUSTIC models, we first generate low-rank $\mathbf{\hat{L}}$ and sparse $\hat{\mathbf{S}}$ targets corresponding to the input $\mathbf{D}$. These targets are used to train the unrolled networks via backpropagation with a suitable loss function. The loss function we choose is the mean squared error (MSE) between the targets $\hat{\mathbf{L}}_i$, $\hat{\mathbf{S}}_i$ and the model predictions $\mathbf{L}_m$, $\mathbf{S}_m$
\begin{equation}
    \mathcal{L}(\mathbf{\theta}) = \frac{1}{2M}\sum_{m=1}^{M} \left(\norm{\mathbf{S}_m - \hat{\mathbf{S}}_m}_F^2 + \norm{\mathbf{L}_m - \hat{\mathbf{L}}_m}_F^2\right).
\end{equation}
We use ISTA to form the targets and will describe the procedure for generating $\hat{\mathbf{L}}$ and $\hat{\mathbf{S}}$ in greater detail in Section~\ref{sec:setup}.

\begin{figure*}[ht]
\centering
    \subfloat[\centering F-Scores for scene A trained on scene A]{{\includegraphics[width=.4\linewidth]{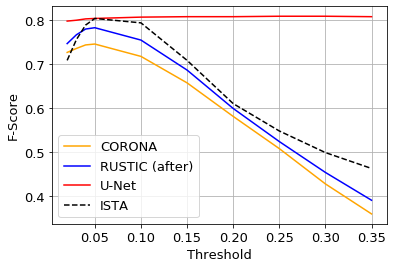} }}%
    \subfloat[\centering F-Scores for scene B trained on scene B]{{\includegraphics[width=.4\linewidth]{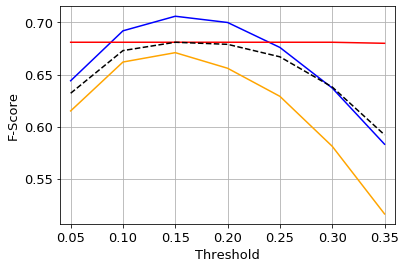} }}%
    
    \subfloat[\centering F-Scores for scene B trained on scene A]{{\includegraphics[width=.4\linewidth]{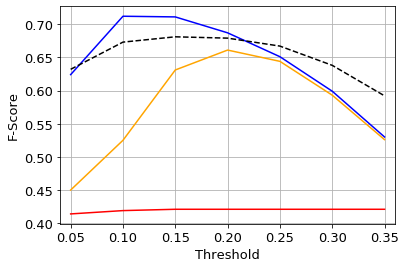} }}%
    \subfloat[\centering F-Scores for for scene A trained on scene B]{{\includegraphics[width=.4\linewidth]{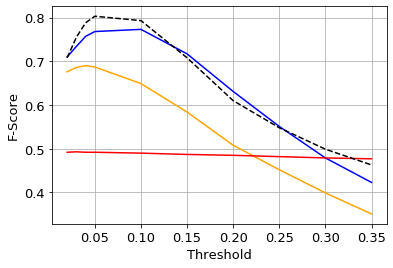} }}%

    \caption{Comparison of RUSTIC, CORONA and U-Net models with their ISTA targets against various thresholds. The means from 5 trials are displayed. (a) and (c) are trained on scene A while (b) and (d) are trained on scene B.}
\label{fig:f_score_general}
\end{figure*}

\begin{figure*}[ht]
\centering
    \subfloat[\centering Input]{{\includegraphics[width=.3\linewidth]{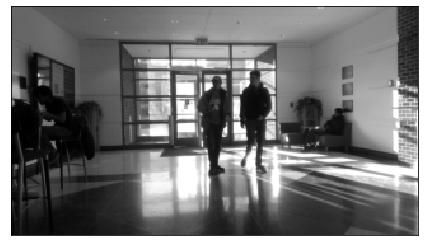} }}%
    \subfloat[\centering Two-layer CORONA ]{{\includegraphics[width=.3\linewidth]{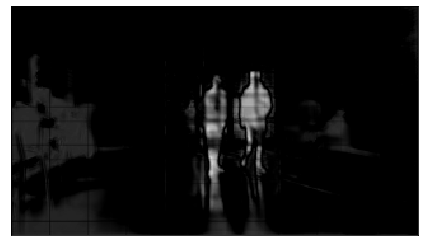} }}%
    \subfloat[\centering Eight-layer CORONA ]{{\includegraphics[width=.3\linewidth]{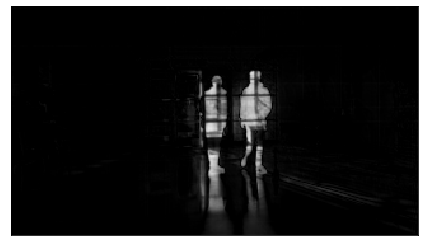} }}%
    
    \subfloat[\centering ISTA ]{{\includegraphics[width=.3\linewidth]{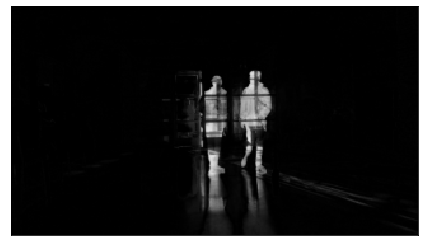} }}%
    \subfloat[\centering Two-layer RUSTIC ]{{\includegraphics[width=.3\linewidth]{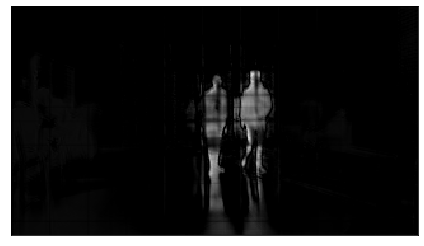} }}%
    \subfloat[\centering Eight-layer RUSTIC ]{{\includegraphics[width=.3\linewidth]{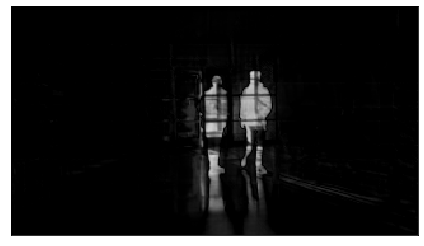} }}%
    \caption{Results from scene B for models trained on scene A. Images (b)-(f) contain the magnitudes of the sparse outputs. The RUSTIC models used in (e) and (f) incorporate the radar \emph{after} the shrinkage operator. The two-layer RUSTIC model with radar does the best job of suppressing the shadows and static humans/furniture on the sides. At eight layers, the models perform relatively similarly. Moreover, the radar data seems less influential in (f) than in (e) since there is a stronger presence of shadows on the right.}%
    \label{fig:lobby_700_all}
\end{figure*}

\section{Experimentation}
\label{sec:experiment}
\begin{figure*}[ht]
\centering
    \subfloat[\centering Input]{{\includegraphics[width=.23\linewidth]{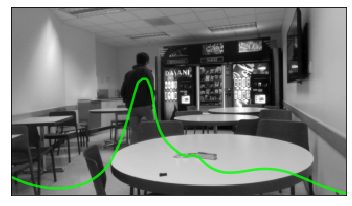} }}%
    \subfloat[\centering ISTA]{{\includegraphics[width=.23\linewidth]{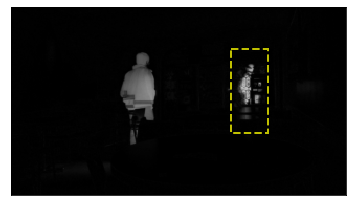} }}%
    \subfloat[\centering Two-layer CORONA]{{\includegraphics[width=.23\linewidth]{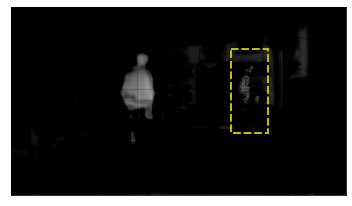} }}%
    \subfloat[\centering Two-layer RUSTIC (\emph{after})]{{\includegraphics[width=.23\linewidth]{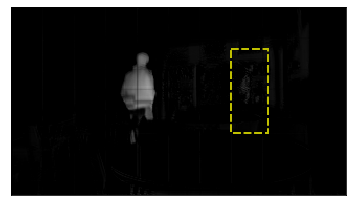} }}%
    \caption{The human pictured above in scene C stands in front of the right vending machine for the majority of the frames (not pictured in this frame) thus causing a ghost human to incorrectly appear in the foreground and background simultaneously as emphasized by the yellow rectangle. Here we see the RUSTIC model in (d) is the only model to properly suppress this error. For context, the radar data is superimposed on (a).}
\label{fig:pantry}
\end{figure*}

\subsection{Setup}
\label{sec:setup}
To evaluate our models, we compare the three variations of RUSTIC depicted in Fig.~\ref{fig:three_models} along with a baseline model without radar (CORONA \cite{corona}) and a standard U-Net \cite{unet}. We use data from RaDICaL, a synchronized FMCW radar, depth, IMU and RGB dataset \cite{radical}. Each sequence contains images downsampled to $180 \times 320$ and transformed to grayscale. Both the camera and radar data are also downsampled in time so that the frame rates are 3 frames per second. As described in \cite{radical}, the radar data was collected using the Texas Instruments IWR1443BOOST and used 4 receiving antennas and 2 transmitting antennas. By exploiting time division multiplexing this configuration yields 8 virtual receiving antennas in the horizontal axis.

We generate the sparse and low-rank components for the targets by solving the RPCA objective using ISTA without radar side-information. We assume identity measurement matrices $\mathbf{I} = \mathbf{H}_1 = \mathbf{H}_2$, thus $\mu=1$, and run it for 400 iterations. By using ISTA instead of RISTA, we avoid having to tune additional hyperparameters ($\mathbf{H}_3$) that control how to properly scale the radar data for fusion with camera data. In order to evaluate models after training as well as the quality of generated ISTA targets, we hand-label binary images with each pixel labeled as either foreground or background. Only desirable foreground components such as moving humans and doors are labeled as foreground.

We experiment with three scenes, labeled A, B, and C, that are all 30 frames long. Only three scenes are used in this work because of the limited amount of synchronized camera/radar data available. 
We believe this amount still provides a sufficient demonstration because these three scenes have distinct background and foreground, contain varying amounts of undesirable foreground like shadows and reflections, and are dissimilar enough to test each model's propensity to overfit. We do not train the model on scene C because the ISTA results are quite poor as shown in Fig.~\ref{fig:pantry}.

For a fair comparison to the unrolled networks, the U-Net model is trained on a single sequence where each image is its own channel. Thus, the input to the U-Net model is of shape $(30, H, W)$. This allows the U-Net to leverage information from the entire sequence instead of single images to predict its output. We also use the generated ISTA targets to train the U-Net; however, it is important to note that the U-Net outputs represent the probability of each pixel being foreground, as is common practice in deep learning BFS models \cite{deep-learning-survey}. This means the U-Net does not perform the same low-rank+sparse separation as the unrolled networks and only predicts the presence of foreground. To generate the U-Net targets, we threshold the magnitude of the sparse components from ISTA to create one-hot probability distribution targets at each pixel. Thus, the U-Net and unrolled models work with the same training data up to this small thresholding modification to train the U-Net. We empirically choose 0.075 for scene A and 0.15 for scene B as the thresholds for $\left|\mathbf{S}_{mhw}\right| \in [0,1]$.


\subsection{Complexity}
\label{sec:complexity}
Like ISTA, the time complexity of each layer in RUSTIC and CORONA is dominated by the SVD and matrix multiplication operations in the SVT operation to update the low rank component. Thus, the time complexity for a $k$-layer network is $\mathcal{O}\left(  k[W^2H^2M + WHM^2 + M^3]  \right)$. The memory requirement is also substantial due to the SVD operation because matrices of size $H^2W^2\times H^2W^2$ and $H^2W^2\times M$ need to be stored during each forward pass through each layer. To make such computation tractable, we process the input image sequence in patches smaller than the image size ($H, W$). During training, each batch consists of one randomly selected patch. Then, for test-time inference we choose a stride length in each dimension less than or equal to the patch size and iterate over the entire image. For cases when the stride length is less than the patch size in either dimension, we take the mean of the regions with overlap.

\subsection{Training Details}
\label{sec:train-details}
We train the models on a single sequence of 30 frames for 50,000 image patches. Our unrolled networks use 2D kernel sizes of $5\times5$ for the first 3 layers, $3\times3$ for all subsequent layers, and length 5 1D convolutions for radar data in all layers. We use the Adam optimizer \cite{adam_opt} and a learning rate of $10^{-3}$ for the first 30,000 patches and $10^{-4}$ for the remaining 20,000. Furthermore, all models are run for five trials with five consistent random seeds shared across all architectures.

To address the high memory requirements of performing SVD, we use patch sizes of $80 \times 80$ for each input to the unrolled networks. To generate full image results, we use a stride length of 30 pixels in each dimension and average predictions results where there is overlap.

\subsection{Results}

\begin{table}
\centering
\caption{The highest achieved F-scores for two and eight-layer models of CORONA and RUSTIC. The means from 5 trials are displayed. The best results for each network depth are bolded and the best results for each row are underlined.}

 \begin{tabular}{c|cc|cc}
   Scene & 2 Layer & 2 Layer & 8 Layer & 8 Layer \\ (Train, Eval) & CORONA & RUSTIC & CORONA & RUSTIC \\ \hline  
  (A,A) & 0.745 & \textbf{0.782}& \underline{\textbf{0.797}} & \underline{\textbf{0.797}} \\
  (A,B) & 0.661 &\underline{\textbf{0.712}} & 0.694 & \textbf{0.702}\\
  (B,A) & 0.690 & \textbf{0.773} & \underline{\textbf{0.808}} & 0.798 \\
  (B,B) & 0.671 & \underline{\textbf{0.706}} & 0.688 & \textbf{0.691}\\
 \end{tabular}
 \label{tab:f_score}
\end{table}

For all figures and tables in this section, RUSTIC refers to the best performing configuration where the radar is used \emph{after} the shrinkage operator unless noted otherwise. Quantitative results for two-layer unrolled models are presented in Fig.~\ref{fig:f_score_general}. We see a clear gap between the two unrolled networks in favor of the RUSTIC architecture. Notably, RUSTIC outperforms the ISTA targets when evaluated on scene B regardless of the scene it is trained on. Thus, RUSTIC provides real-time computation and superior performance on this scene. Figure~\ref{fig:lobby_700_all} shows example sparse outputs for scene B from models trained on scene A. For the displayed image, only the two walking humans in the middle and the closing door are labeled as true foreground. From the figure, we see that the two-layer RUSTIC model detects the true foreground similarly to ISTA and CORONA but does a much better job of suppressing the shadows to the right of the humans.  This suggests that the side-information from the radar successfully disagrees with the camera data and yields a more precise foreground. When we increase the network depth to eight layers, we see the results from RUSTIC and CORONA become more similar as the eight-layer RUSTIC model includes more shadows in its foreground. This phenomenon indicates that the radar is used less in deeper models. We also note that because no elevation data is collected by the radar, there are no cues to reduce the reflections below the humans. With access to elevation data, we would expect radar side information to further alleviate this failure mode for camera data.

We also note in Fig.~\ref{fig:f_score_general} the expected overfitting of the U-Net to its training data while poorly generalizing to unseen scenes. The U-Net is only able to match the maximum performance of the ISTA method in \ref{fig:f_score_general}a and \ref{fig:f_score_general}b because the U-Net targets are generated directly from thresholding the ISTA results. Lastly, as seen in Fig.~\ref{fig:lobby_700_all}e and corroborated by Fig.~\ref{fig:f_score_general}c and Table \ref{tab:f_score}, two-layer RUSTIC models perform the best qualitatively and according to F-score on scenes with high amounts of shadow. This is notable because such a shallow model (1) takes more influence from the radar and (2) is faster than its deeper counterparts. We argue for this first point in particular since the unrolled networks with and without radar perform closely with the deeper eight-layer architecture. Supporting numerical results for both two and eight-layer models are presented in Table \ref{tab:f_score}. For the following comparisons between RUSTIC and CORONA, we will use models with two layers.

We also compare the number of parameters and computation time of each method in Table \ref{tab:params-and-time}. We see the expected dramatic gap between the unrolled models and the standard U-Net in number of parameters. Lastly, we observe that both unrolled models support real-time computation unlike their iterative ISTA counterpart.

\begin{table}
\centering
\caption{Number of trainable parameters and average inference time for each method using setup from Section~\ref{sec:setup}. ISTA is run on an Intel\textregistered{}  Core\textsuperscript{\tiny{TM}} i7 7th Gen processor while the networks are run on an NVIDIA GTX 1070 GPU.}
 \begin{tabular}{cccc}
  Method & \# of Parameters & Inference Time (s) & Mean FPS \\ \hline  
  ISTA/RISTA & 0 & 20.612 $\pm$ 0.099 & 1.46\\
  2-Layer CORONA & 316 & 1.157 $\pm$ 0.020 & 25.93\\
  2-Layer RUSTIC & 328 & 1.150 $\pm$ 0.002 & 26.08\\
  8-Layer CORONA & 784 & 2.350 $\pm$ 0.005 & 12.76\\
  8-Layer RUSTIC & 832 & 2.368 $\pm$ 0.014 & 12.67\\
  U-Net & 13,412,766 & 0.399 $\pm$ 0.003 & 75.23\\
 \end{tabular}
 \label{tab:params-and-time}
\end{table}



 \subsubsection{Sleeping Foreground}
 \label{sec:sleep}
 In scene C, we see dramatic qualitative results as we address a situation with \emph{sleeping foreground}. Sleeping foreground refers to the scenario where a foreground object, in this case a human, behaves as clear moving foreground for some frames and then remains relatively still for a large portion of the remaining frames. For the sequence depicted in Fig.~\ref{fig:pantry}, the human stands in front of the vending machine on the right for last 18/30 frames causing both the ISTA algorithm and the two-layer CORONA network to mistakenly absorb them into the low-rank background. As a result, when the person isn't standing at the vending machine like the frame shown in the figure, the sparse foreground must compensate by outputting a \emph{ghost human} in the foreground component. Yet, for the same image, we see that RUSTIC is able to much more effectively suppress the appearance of this ghost human. As mentioned earlier, because deeper networks rely more on the camera data and less on the radar data, the eight-layer models both with and without radar perform poorly and are unable to suppress this instance of incorrect foreground.

\subsubsection{Dealing with Radar False Positives}
\begin{figure}%
    \centering
    \subfloat[\centering Input image with radar return]{{\includegraphics[height=2.38cm, trim=0 5 0 0]{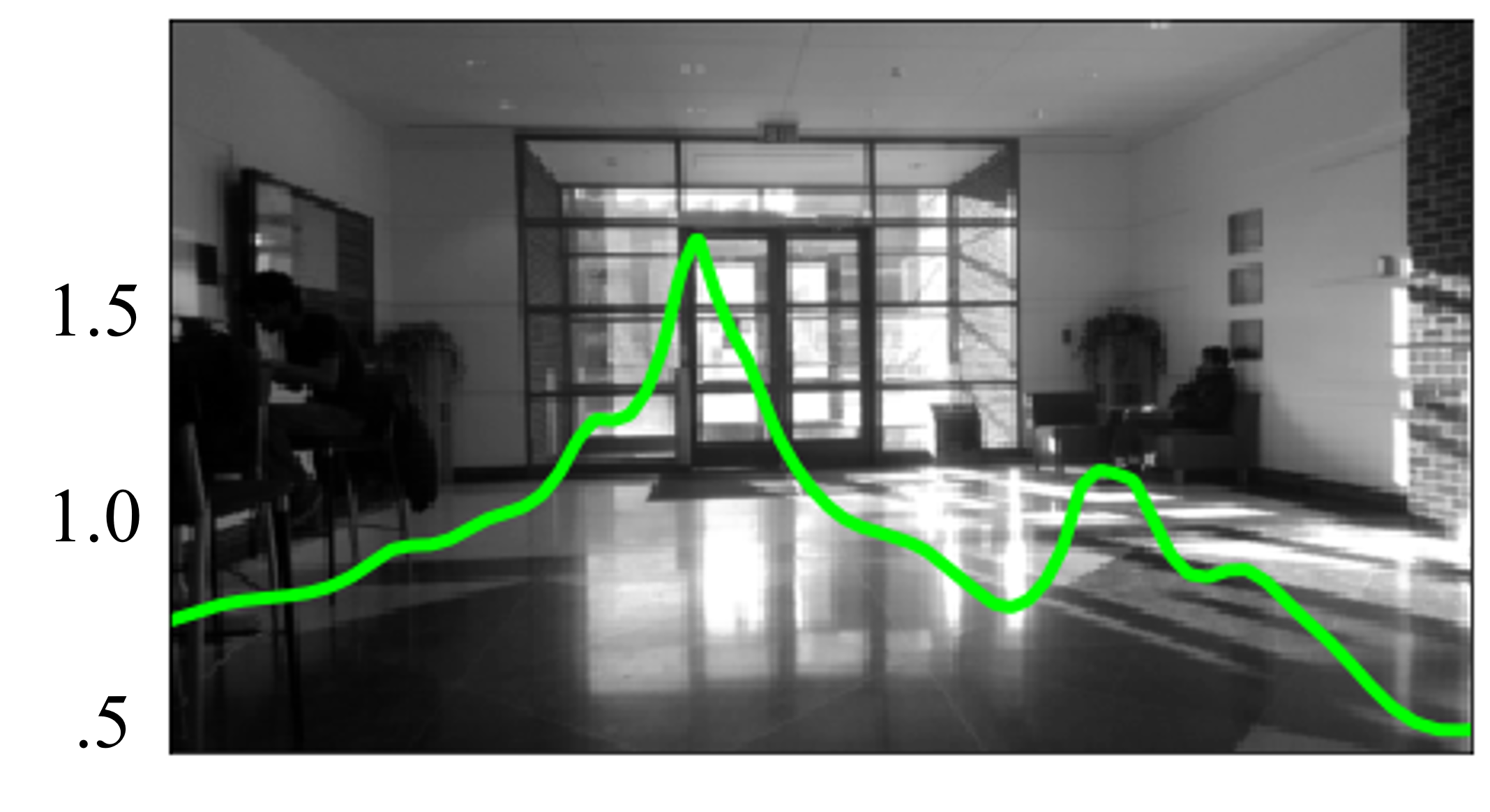} }}%
    \subfloat[\centering RUSTIC sparse output ]{{\includegraphics[height=2.4cm]{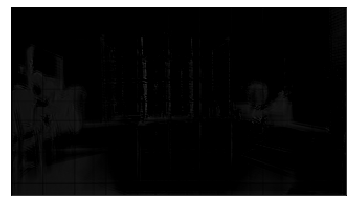} }}%
    
    \caption{In (a) we see a camera image with no visible foreground while the radar return after clutter suppression is overlayed on top. Because there is no visible foreground, (b) should be entirely zero (black), which is nearly the case.
    For context, the walking humans earlier in the frame had returns with magnitudes between 1 and 2.75.}%
    \label{fig:disagreement}%
\end{figure}

Many instances of undesirable foreground in the camera data can be thought of as false positives that are suppressed by the incorporation of the radar data. As mentioned in Fig.~\ref{fig:venn}, the opposite may also occur when the radar mistakenly detects motion when there is none to be seen in the corresponding image. This could be due multipath or motion that is occluded to the camera i.e. behind a wall or inside an opaque container. In one instance, shown in Fig.~\ref{fig:disagreement}, there are two visible peaks in the radar return after clutter suppression as shown overlayed on the image. For context, the walking humans in this sequence have radar returns with magnitudes ranging between 1-2.75. Because there actually is no visible foreground, the sparse component in \ref{fig:disagreement}b should be entirely zero (black). Despite this misleading radar return, RUSTIC correctly suppresses the foreground component thus demonstrating the model's ability to suppress false positives from either sensing modality.

\subsubsection{Comparison of the Models with Radar}

\begin{figure}[h]
\centering
    \subfloat[\centering F-Scores for scene A trained on scene A]{{\includegraphics[width=.75\linewidth]{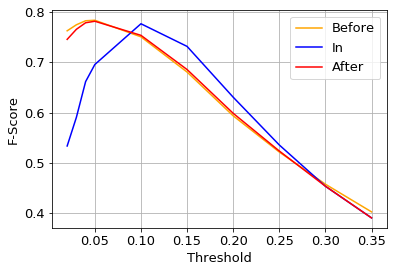} }}%
    
    \subfloat[\centering F-Scores for scene B trained on scene A]{{\includegraphics[width=.75\linewidth]{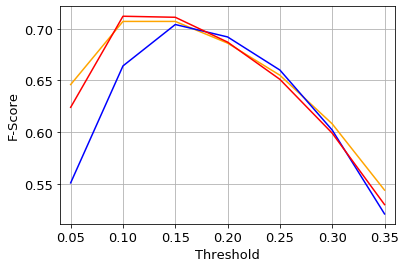} }}%
    \caption{A comparison of the \emph{in}, \emph{before}, and \emph{after} two-layer RUSTIC models. The means from 5 trials are displayed.}
\label{fig:f_score_radar}
\end{figure}

Thus far, all results generated using RUSTIC incorporate the radar \emph{after} the shrinkage operator. In Fig.~\ref{fig:f_score_radar}, we offer a comparison of the three different models described in (\ref{eqn:in_description}), (\ref{eqn:before_description}), (\ref{eqn:after_description}) and depicted in Fig.~\ref{fig:three_models}. All three models were trained on scene A and tested on scene B.

In Fig.~\ref{fig:f_score_radar}a, we see nearly identical performance for the \emph{before} and \emph{after} models. We also see a relatively high peak in training performance for the \emph{in} model but at a higher threshold. This suggests the \emph{in} model produces sparse foregrounds with lower precision.

Furthermore, during our experimentation, we noticed that certain training runs for the \emph{in} model resulted in models that incorrectly predict the sparse components as all zeros. Thus, the results in Fig.~\ref{fig:f_score_radar} only include the models that do not suffer from this instability. We address this issue during training in the following subsection.

\subsubsection{Ablation Study with Cosine Similarity Loss}
\label{sec:cos}

As mentioned above, the models that incorporate the radar \emph{in} the shrinkage operator are prone to local minima where the low rank outputs are learned correctly and the sparse components give all zeros. To rectify this, we add to the loss function a scaled cosine similarity term between the $l_1$ norm of the columns in $\mathbf{S}_m$ and $\mathbf{R}_m$
\begin{align} \alpha
    \frac{\langle \sum_{h=1}^H |\mathbf{S}_{mh}|,\mathbf{R}_m\rangle}{\norm{\sum_{h=1}^H |\mathbf{S}_{mh}|}_2\norm{\mathbf{R}_m}_2}
\end{align}
where $\alpha\in[0, 1]$. This loss term assumes the amount of sparse foreground in a given column is proportional to its radar return. The absolute value ensures that negative and positive foreground intensities are treated identically.
In our experiments, we empirically set $\alpha=10^{-3}$ to appropriately balance the MSE loss. With this choice, we observe that all runs avoid any local minima and the performance is otherwise unaffected for better or worse.

    



\section{Conclusion}
\label{sec:conclusion}

In this work, we present a number of contributions to BFS. First,  we motivated the incorporation of radar data into the RPCA objective and introduced an associated iterative solver called RISTA. We then unrolled our iterative algorithm into our RUSTIC model and tested our approach in the unsupervised setting where no ground-truth is available. We found that RUSTIC provided real-time computation without sacrificing the performance from the associated iterative solver. While we do notice some convergence issues with incorporating the radar \emph{in} the shrinkage operator, we mitigated this issue with the addition of a cosine similarity loss term during training.


We also demonstrated strong performance in scenarios when the camera data and radar disagree. We showed that the two-layer RUSTIC network is able to effectively suppress shadows and ignore sleeping foreground objects. Moreover, in the case with improper radar returns, we saw that the sparse output did not contain strong unwanted foreground when the radar incorrectly encouraged otherwise.

Finally, we saw in Fig.~\ref{fig:lobby_700_all} that deeper models performed more closely to ISTA and seemed to incorporate radar information less. This phenomenon was most pronounced for quantitative results with two-layer unrolled networks as RUSTIC clearly outperformed CORONA. This provides evidence that deeper unrolled models may not always be best, especially when additional modalities are available.


While this work does demonstrate the efficacy of using radar reflections at a given bearing for BFS, much of the radar information remains unused. For example, with priors on the types of targets that may be observed in a scene, the radar's range and magnitude information could provide insight on how much area in pixels the targets might occupy. Furthermore, more processing on the velocity information could also prove useful in extracting desirable foreground. This velocity information could be incorporated into a tracking scheme that yields more reliable and consistent foreground concepts.
Moreover, in some cases users may only be interested in viewing foreground targets that fall within a certain range of Doppler velocities. This might be useful in distinguishing between moving vehicles and walking pedestrians.

Finally, we believe that the sensor fusion methods presented in this work are not limited to radar. Other sensors such as sonar and lidar likely could also be used as long as the processing can eliminate static clutter reliably.





%
\bibliographystyle{./bibliography/IEEEtran}
\bibliography{./bibliography/IEEEbib}

\begin{thebibliography}{10}
\providecommand{\url}[1]{#1}
\csname url@samestyle\endcsname
\providecommand{\newblock}{\relax}
\providecommand{\bibinfo}[2]{#2}
\providecommand{\BIBentrySTDinterwordspacing}{\spaceskip=0pt\relax}
\providecommand{\BIBentryALTinterwordstretchfactor}{4}
\providecommand{\BIBentryALTinterwordspacing}{\spaceskip=\fontdimen2\font plus
\BIBentryALTinterwordstretchfactor\fontdimen3\font minus
  \fontdimen4\font\relax}
\providecommand{\BIBforeignlanguage}[2]{{%
\expandafter\ifx\csname l@#1\endcsname\relax
\typeout{** WARNING: IEEEtran.bst: No hyphenation pattern has been}%
\typeout{** loaded for the language `#1'. Using the pattern for}%
\typeout{** the default language instead.}%
\else
\language=\csname l@#1\endcsname
\fi
#2}}
\providecommand{\BIBdecl}{\relax}
\BIBdecl

\bibitem{bfs-applications}
\BIBentryALTinterwordspacing
B.~Garcia-Garcia, T.~Bouwmans, and A.~J. {Rosales Silva}, ``Background
  subtraction in real applications: Challenges, current models and future
  directions,'' \emph{Computer Science Review}, vol.~35, p. 100204, 2020.
  [Online]. Available:
  \url{https://www.sciencedirect.com/science/article/pii/S1574013718303101}
\BIBentrySTDinterwordspacing

\bibitem{rpca-survey}
\BIBentryALTinterwordspacing
T.~Bouwmans and E.~H. Zahzah, ``Robust pca via principal component pursuit: A
  review for a comparative evaluation in video surveillance,'' \emph{Computer
  Vision and Image Understanding}, vol. 122, pp. 22--34, 2014. [Online].
  Available:
  \url{https://www.sciencedirect.com/science/article/pii/S1077314213002294}
\BIBentrySTDinterwordspacing

\bibitem{deep-learning-survey}
\BIBentryALTinterwordspacing
T.~Bouwmans, S.~Javed, M.~Sultana, and S.~K. Jung, ``Deep neural network
  concepts for background subtraction:a systematic review and comparative
  evaluation,'' \emph{Neural Networks}, vol. 117, pp. 8--66, 2019. [Online].
  Available:
  \url{https://www.sciencedirect.com/science/article/pii/S0893608019301303}
\BIBentrySTDinterwordspacing

\bibitem{classical-survey}
\BIBentryALTinterwordspacing
T.~Bouwmans, ``Traditional and recent approaches in background modeling for
  foreground detection: An overview,'' \emph{Computer Science Review}, vol.
  11-12, pp. 31--66, 2014. [Online]. Available:
  \url{https://www.sciencedirect.com/science/article/pii/S1574013714000033}
\BIBentrySTDinterwordspacing

\bibitem{rpca}
E.~J. Cand{\`e}s, X.~Li, Y.~Ma, and J.~Wright, ``Robust principal component
  analysis?'' \emph{Journal of the ACM (JACM)}, vol.~58, no.~3, pp. 1--37,
  2011.

\bibitem{faster-rpca}
Z.~Lin, A.~Ganesh, J.~Wright, L.~Wu, M.~Chen, and Y.~Ma, ``Fast convex
  optimization algorithms for exact recovery of a corrupted low-rank matrix,''
  \emph{Coordinated Science Laboratory Report no. UILU-ENG-09-2214, DC-246},
  2009.

\bibitem{realtime-rpca}
C.~Qiu and N.~Vaswani, ``Real-time robust principal components' pursuit,'' in
  \emph{2010 48th Annual Allerton Conference on Communication, Control, and
  Computing (Allerton)}, 2010, pp. 591--598.

\bibitem{orpca}
J.~Feng, H.~Xu, and S.~Yan, ``Online robust pca via stochastic optimization,''
  in \emph{Advances in Neural Information Processing Systems}, C.~J.~C. Burges,
  L.~Bottou, M.~Welling, Z.~Ghahramani, and K.~Q. Weinberger, Eds.,
  vol.~26.\hskip 1em plus 0.5em minus 0.4em\relax Curran Associates, Inc.,
  2013.

\bibitem{cdnet14}
Y.~Wang, P.-M. Jodoin, F.~Porikli, J.~Konrad, Y.~Benezeth, and P.~Ishwar,
  ``Cdnet 2014: An expanded change detection benchmark dataset,'' in \emph{2014
  IEEE Conference on Computer Vision and Pattern Recognition Workshops}, 2014,
  pp. 393--400.

\bibitem{SBI2015}
L.~Maddalena and A.~Petrosino, ``Towards benchmarking scene background
  initialization,'' in \emph{New Trends in Image Analysis and Processing --
  ICIAP 2015 Workshops}, V.~Murino, E.~Puppo, D.~Sona, M.~Cristani, and
  C.~Sansone, Eds., 2015.

\bibitem{fgsegnet}
\BIBentryALTinterwordspacing
L.~A. Lim and H.~Y. Keles, ``{Learning multi-scale features for foreground
  segmentation},'' \emph{Pattern Analysis and Applications}, vol.~23, no.~3,
  pp. 1369--1380, 2020. [Online]. Available:
  \url{https://doi.org/10.1007/s10044-019-00845-9}
\BIBentrySTDinterwordspacing

\bibitem{cascade-cnn}
\BIBentryALTinterwordspacing
Y.~Wang, Z.~Luo, and P.-M. Jodoin, ``Interactive deep learning method for
  segmenting moving objects,'' \emph{Pattern Recognition Letters}, vol.~96, pp.
  66--75, 2017, scene Background Modeling and Initialization. [Online].
  Available:
  \url{https://www.sciencedirect.com/science/article/pii/S0167865516302471}
\BIBentrySTDinterwordspacing

\bibitem{deep-context-prediction}
\BIBentryALTinterwordspacing
M.~Sultana, A.~Mahmood, S.~Javed, and S.~K. Jung, ``{Unsupervised deep context
  prediction for background estimation and foreground segmentation},''
  \emph{Machine Vision and Applications}, vol.~30, no.~3, pp. 375--395, 2019.
  [Online]. Available: \url{https://doi.org/10.1007/s00138-018-0993-0}
\BIBentrySTDinterwordspacing

\bibitem{first_unroll}
K.~Gregor and Y.~LeCun, ``Learning fast approximations of sparse coding,'' in
  \emph{Proceedings of the 27th International Conference on Machine Learning},
  2010, pp. 399--406.

\bibitem{monga2021algorithm}
V.~Monga, Y.~Li, and Y.~C. Eldar, ``Algorithm unrolling: Interpretable,
  efficient deep learning for signal and image processing,'' \emph{IEEE Signal
  Processing Magazine}, vol.~38, no.~2, pp. 18--44, 2021.

\bibitem{image_deblur}
Y.~Li, M.~Tofighi, J.~Geng, V.~Monga, and Y.~C. Eldar, ``Efficient and
  interpretable deep blind image deblurring via algorithm unrolling,''
  \emph{IEEE Transactions on Computational Imaging}, vol.~6, pp. 666--681,
  2020.

\bibitem{phase_retrieval}
R.~Hyder, Z.~Cai, and M.~S. Asif, ``Solving phase retrieval with a learned
  reference,'' in \emph{European Conference on Computer Vision}, 2020, pp.
  425--441.

\bibitem{viterbi_net}
N.~Shlezinger, N.~Farsad, Y.~C. Eldar, and A.~J. Goldsmith, ``Viterbinet: A
  deep learning based {V}iterbi algorithm for symbol detection,'' \emph{IEEE
  Transactions on Wireless Communications}, vol.~19, no.~5, pp. 3319--3331,
  2020.

\bibitem{corona}
O.~Solomon, R.~Cohen, Y.~Zhang, Y.~Yang, Q.~He, J.~Luo, R.~J. van Sloun, and
  Y.~C. Eldar, ``Deep unfolded robust {PCA} with application to clutter
  suppression in ultrasound,'' \emph{IEEE Transactions on Medical Imaging},
  vol.~39, no.~4, pp. 1051--1063, 2019.

\bibitem{radical}
T.~Y. Lim, S.~Markowitz, and M.~N. Do, ``Radical: A synchronized {FMCW} radar,
  depth, {IMU} and {RGB} camera data dataset with low-level {FMCW} radar
  signals,'' \emph{IEEE Journal of Selected Topics in Signal Processing}, 2021.

\bibitem{unet}
O.~Ronneberger, P.~Fischer, and T.~Brox, ``U-net: Convolutional networks for
  biomedical image segmentation,'' in \emph{International Conference on Medical
  Image Computing and Computer-Assisted Intervention}, 2015, pp. 234--241.

\bibitem{mvdr}
J.~Capon, ``High-resolution frequency-wavenumber spectrum analysis,''
  \emph{Proceedings of the IEEE}, vol.~57, no.~8, pp. 1408--1418, 1969.

\bibitem{sprechmann2015learning}
P.~Sprechmann, A.~M. Bronstein, and G.~Sapiro, ``Learning efficient sparse and
  low rank models,'' \emph{IEEE Transactions on Pattern Analysis and Machine
  Intelligence}, vol.~37, no.~9, pp. 1821--1833, 2015.

\bibitem{adam_opt}
\BIBentryALTinterwordspacing
D.~P. Kingma and J.~Ba, ``Adam: {A} method for stochastic optimization,'' in
  \emph{3rd International Conference on Learning Representations, {ICLR} 2015,
  San Diego, CA, USA, May 7-9, 2015, Conference Track Proceedings}, Y.~Bengio
  and Y.~LeCun, Eds., 2015. [Online]. Available:
  \url{http://arxiv.org/abs/1412.6980}
\BIBentrySTDinterwordspacing

\end{thebibliography}

%

 \begin{IEEEbiography}[{\includegraphics[width=1in,height=1.25in,clip,keepaspectratio]{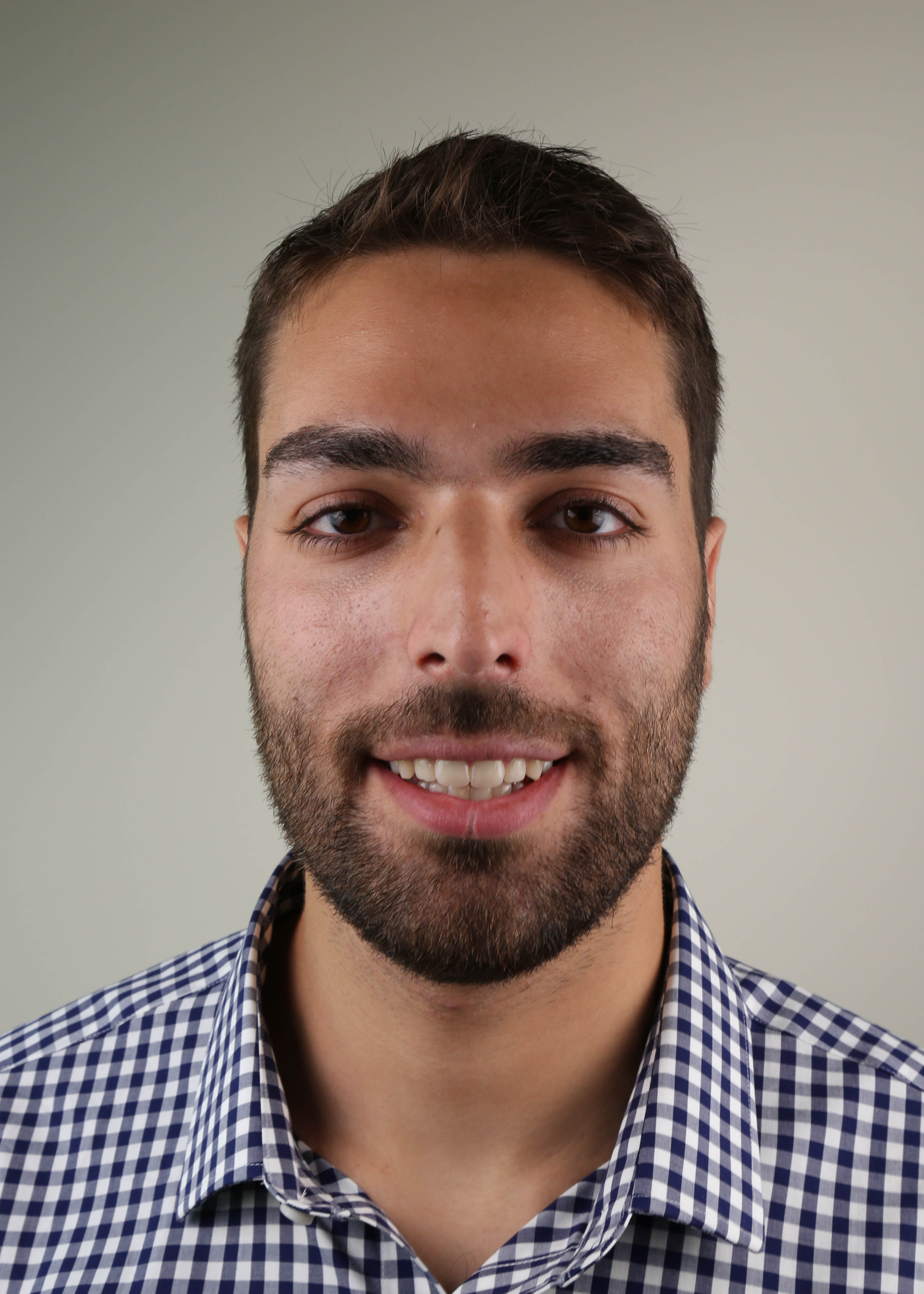}}]{Spencer Markowitz}
  is a graduate student in the Electrical and Computer Engineering Department at the University of Illinois at Urbana-Champaign. He also received his B.S. in electrical engineering at the University of Illinois. His primary research interests include FMCW radar, computer vision, object tracking, and deep learning. 
  \end{IEEEbiography}

\enlargethispage{-9.5cm}

\begin{IEEEbiography}[{\includegraphics[width=1in,height=1.25in,clip,keepaspectratio]{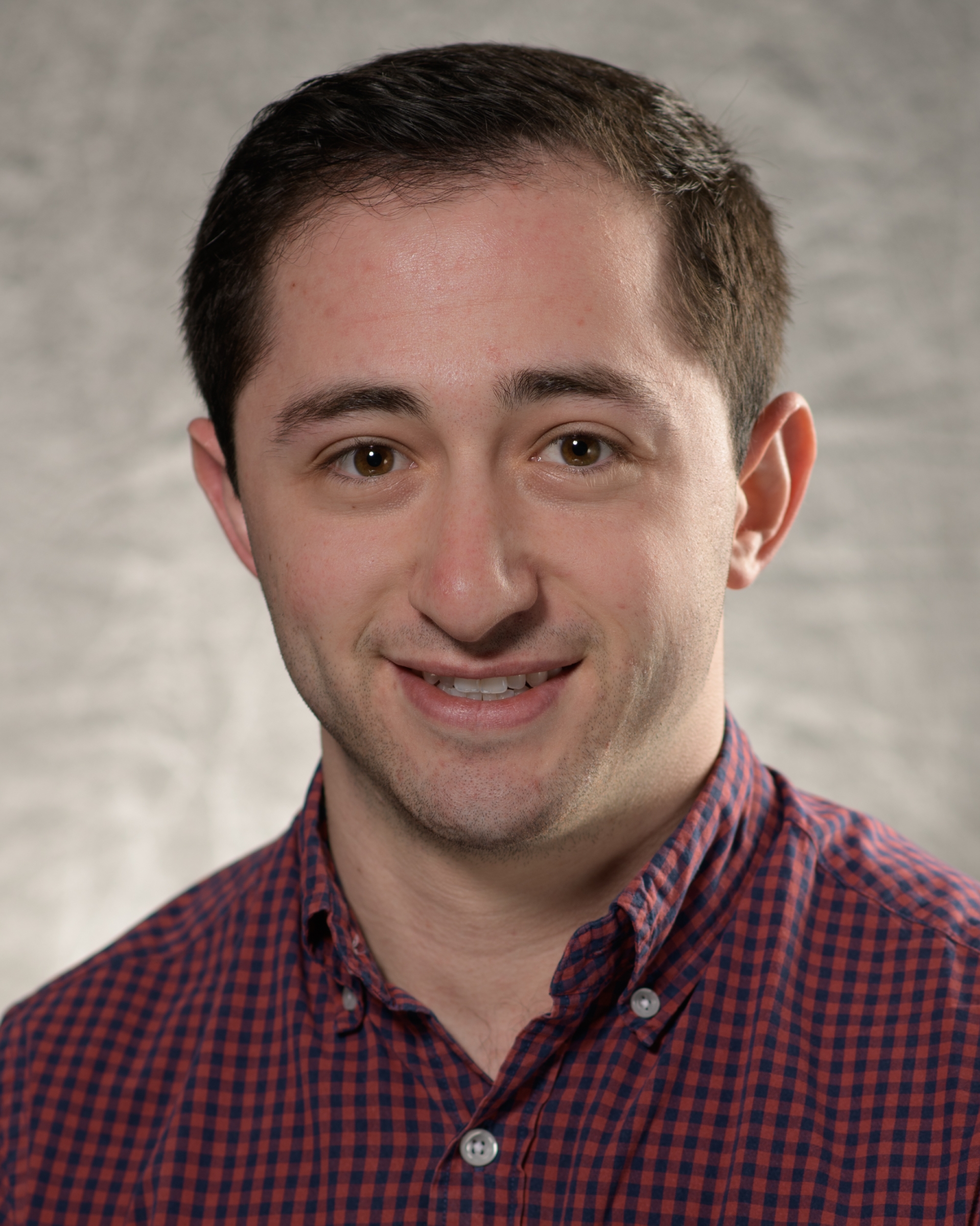}}]{Corey Snyder}
is a graduate student in the Electrical and Computer Engineering Department at the University of Illinois Urbana-Champaign. He received his B.S. in electrical engineering in 2018 and M.S. in electrical and computer engineering in 2020, both from the University of Illinois Urbana-Champaign. His research interests include semi-supervised, weakly supervised, and unsupervised learning for computer vision. \end{IEEEbiography}

\begin{IEEEbiography}[{\includegraphics[width=1in,height=1.25in,clip,keepaspectratio]{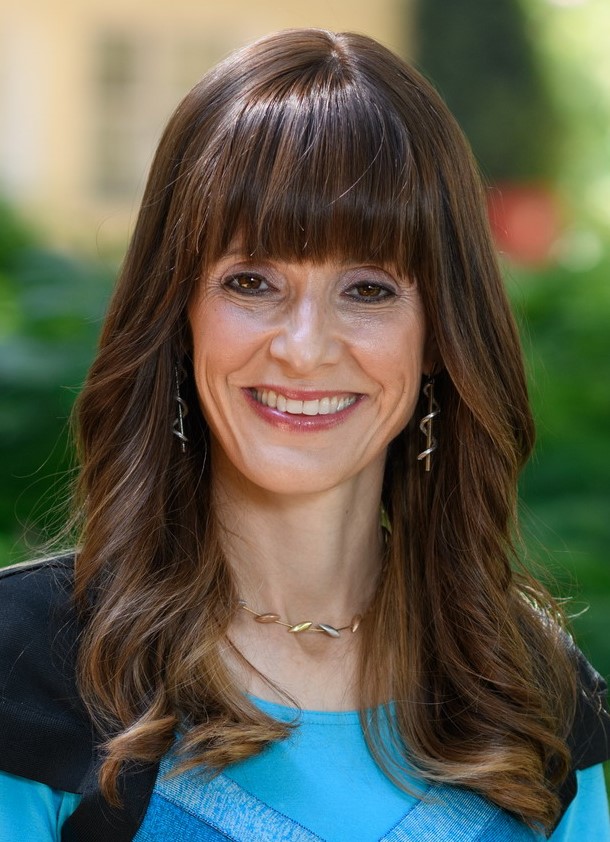}}]{Yonina C. Eldar}
Yonina C. Eldar (S’98–M’02–SM’07-F'12) received the B.Sc. degree in Physics in 1995 and the B.Sc. degree in Electrical Engineering in 1996 both from Tel-Aviv University (TAU), Tel-Aviv, Israel, and the Ph.D. degree in Electrical Engineering and Computer Science in 2002 from the Massachusetts Institute of Technology (MIT), Cambridge.

She is currently a Professor in the Department of Mathematics and Computer Science, Weizmann Institute of Science, Rehovot, Israel. She was previously a Professor in the Department of Electrical Engineering at the Technion, where she held the Edwards Chair in Engineering. She is also a Visiting Professor at MIT, a Visiting Scientist at the Broad Institute, and an Adjunct Professor at Duke University and was a Visiting Professor at Stanford. She is a member of the Israel Academy of Sciences and Humanities (elected 2017), an IEEE Fellow and a EURASIP Fellow. Her research interests are in the broad areas of statistical signal processing, sampling theory and compressed sensing, learning and optimization methods, and their applications to biology, medical imaging and optics.

Dr. Eldar has received many awards for excellence in research and teaching, including the  IEEE Signal Processing Society Technical Achievement Award (2013), the IEEE/AESS Fred Nathanson Memorial Radar Award (2014), and the IEEE Kiyo Tomiyasu Award (2016). She was a Horev Fellow of the Leaders in Science and Technology program at the Technion and an Alon Fellow. She received the Michael Bruno Memorial Award from the Rothschild Foundation, the Weizmann Prize for Exact Sciences, the Wolf Foundation Krill Prize for Excellence in Scientific Research, the Henry Taub Prize for Excellence in Research (twice), the Hershel Rich Innovation Award (three times), the Award for Women with Distinguished Contributions, the Andre and Bella Meyer Lectureship, the Career Development Chair at the Technion, the Muriel \& David Jacknow Award for Excellence in Teaching, and the Technion’s Award for Excellence in Teaching (two times).  She received several best paper awards and best demo awards together with her research students and colleagues including the SIAM outstanding Paper Prize, the UFFC Outstanding Paper Award, the Signal Processing Society Best Paper Award and the IET Circuits, Devices and Systems Premium Award, was selected as one of the 50 most influential women in Israel and in Asia, and is a highly cited researcher.

She was a member of the Young Israel Academy of Science and Humanities and the Israel Committee for Higher Education. She is the Editor in Chief of Foundations and Trends in Signal Processing, a member of the IEEE Sensor Array and Multichannel Technical Committee and serves on several other IEEE committees. In the past, she was a Signal Processing Society Distinguished Lecturer, member of the IEEE Signal Processing Theory and Methods and Bio Imaging Signal Processing technical committees, and served as an associate editor for the IEEE Transactions On Signal Processing, the EURASIP Journal of Signal Processing, the SIAM Journal on Matrix Analysis and Applications, and the SIAM Journal on Imaging Sciences. She was Co-Chair and Technical Co-Chair of several international conferences and workshops. She is author of the book "Sampling Theory: Beyond Bandlimited Systems" and co-author of four other books published by Cambridge University Press.
\end{IEEEbiography}

\begin{IEEEbiography}[{\includegraphics[width=1in,height=1.25in,clip,keepaspectratio]{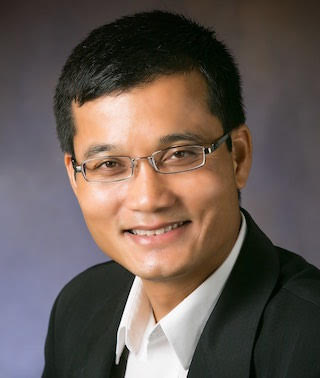}}]{Minh N. Do}
(M'01, SM'07, F'14) was born in Vietnam in 1974.  He received the B.Eng. degree in computer engineering from the University of Canberra, Australia, in 1997, and the Dr.Sci. degree in communication systems from the Swiss Federal Institute of Technology Lausanne (EPFL), Switzerland, in 2001.

Since 2002, he has been on the faculty at the University of Illinois at Urbana-Champaign (UIUC), where he is currently the Thomas and Margaret Huang Endowed Professor in Signal Processing \& Data Science in the Department of Electrical and Computer Engineering, and hold affiliate appointments with the Coordinated Science Laboratory, the Beckman Institute for Advanced Science and Technology, Department of Bioengineering, and the Department of Computer Science.  In 2020-2021, he is on leave from UIUC to serve as the Vice Provost of VinUniversity in Vietnam. 

He received a Silver Medal from the 32nd International Mathematical Olympiad in 1991, University Medal from the University of Canberra in 1997, Doctorate Award from the EPFL in 2001, CAREER Award from the National Science Foundation in 2003,  Xerox Award for Faculty Research from UIUC in 2007, and Young Author Best Paper Award from IEEE in 2008.  He was an Associate Editor of the IEEE Transactions on Image Processing, and a member of several IEEE Technical Committees on Signal Processing.  He was elected as an IEEE Fellow in 2014 for his contributions to image representation and computational imaging.  He has contributed to several tech-transfer efforts, including as a co-founder and CTO of Personify and Chief Scientist of Misfit.\end{IEEEbiography}

\clearpage
\onecolumn
\appendices
    
    

    


\section{Sample Images}
\begin{figure*}[h]
\centering
    \subfloat[\centering Frame 0]{{\includegraphics[width=.3\linewidth]{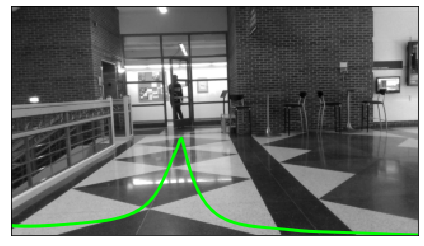} }}%
    \subfloat[\centering Frame 10]{{\includegraphics[width=.3\linewidth]{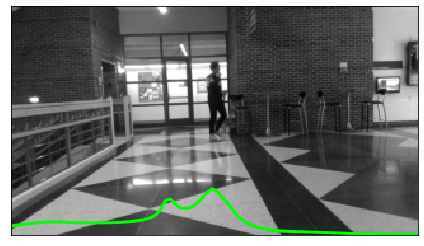} }}%
    \subfloat[\centering Frame 20]{{\includegraphics[width=.3\linewidth]{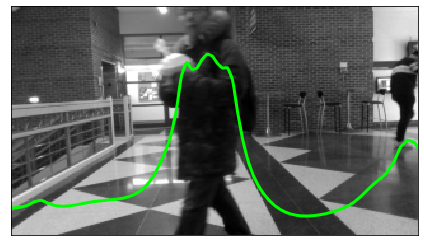} }}%
    
    
    \caption{Sample images and radar data from scene A}
\label{fig:scene_a}
\end{figure*}

\begin{figure*}[h]
\centering
    \subfloat[\centering Frame 0]{{\includegraphics[width=.3\linewidth]{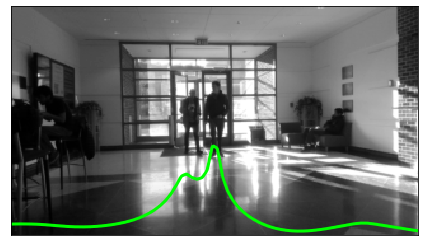} }}%
    \subfloat[\centering Frame 10]{{\includegraphics[width=.3\linewidth]{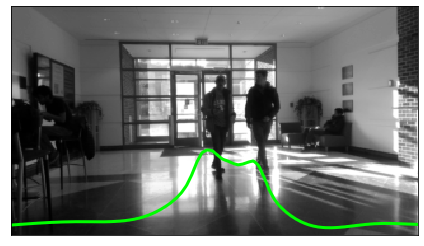} }}%
    \subfloat[\centering Frame 20]{{\includegraphics[width=.3\linewidth]{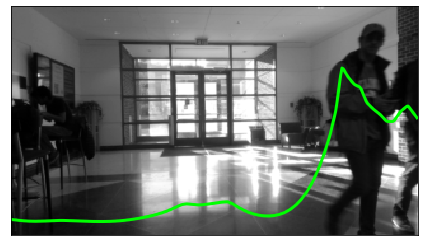} }}%
    
    
    \caption{Sample images and radar data from scene B}
\label{fig:scene_b}
\end{figure*}

\begin{figure*}[h]
\centering
    \subfloat[\centering Frame 0]{{\includegraphics[width=.3\linewidth]{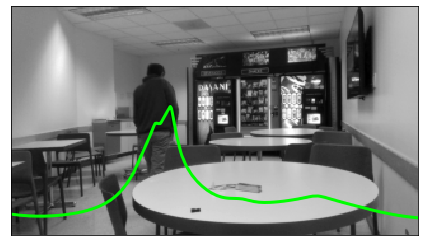} }}%
    \subfloat[\centering Frame 10]{{\includegraphics[width=.3\linewidth]{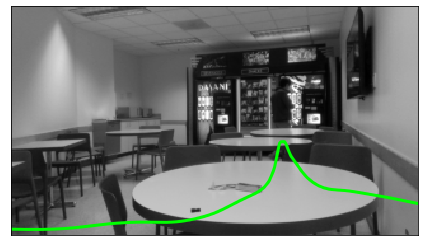} }}%
    \subfloat[\centering Frame 20]{{\includegraphics[width=.3\linewidth]{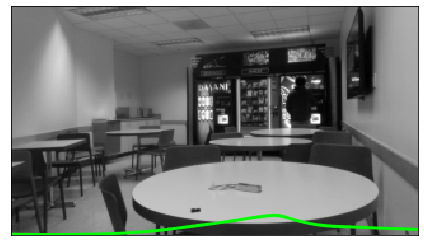} }}%
    
    
    \caption{Sample images and radar data from scene C}
\label{fig:scene_c}
\end{figure*}






\end{document}